\documentclass[journal=jctcce, manuscript=article, layout=onecolumn]{achemso}

\makeatletter\setkeys{acs}{doi=true}\makeatother
\usepackage{amsmath,amssymb}
\usepackage{graphicx}
\usepackage{booktabs}
\usepackage{array}
\usepackage{xcolor}
\usepackage{ifthen}
\usepackage[hidelinks]{hyperref}
\usepackage{xurl}
\usepackage{afterpage}
\usepackage{placeins}
\usepackage{pdflscape}

\usepackage{longtable}
\usepackage[numbers,super,sort&compress]{natbib}

\newcommand{\statusblock}{  \ifthenelse{\equal{\docstatus}{submission}}{}{    \begin{center}\logostrip\end{center}\vspace{0.3em}    \ifthenelse{\equal{\docstatus}{draft}}{\draftbanner}{}  }}

\author{Pavlo O. Dral}
\email{dral@xmu.edu.cn}
\affiliation[XMU]{State Key Laboratory of Physical Chemistry of Solid Surfaces, College of Chemistry and Chemical Engineering, and Fujian Provincial Key Laboratory of Theoretical and Computational Chemistry, Xiamen University, Xiamen, Fujian 361005, China}
\alsoaffiliation[UMK]
{Institute of Physics, Faculty of Physics, Astronomy, and Informatics, Nicolaus Copernicus University in Toru\'n, ul. Grudziadzka 5, 87-100 Toru\'n, Poland}
\alsoaffiliation[UMK2]
{Institute of Advanced Studies, Nicolaus Copernicus University in Toru\'n, ul. Wile\'nska 4, 87-100 Toru\'n, Poland}
\alsoaffiliation[Aitomistic]
{Aitomistic, Shenzhen 518000, China}

\author{Hassan Nawaz}
\affiliation[XMU]{State Key Laboratory of Physical Chemistry of Solid Surfaces, College of Chemistry and Chemical Engineering, and Fujian Provincial Key Laboratory of Theoretical and Computational Chemistry, Xiamen University, Xiamen, Fujian 361005, China}

\author{Arif Ullah}
\email{arif@ahu.edu.cn}
\affiliation{School of Physics, Anhui University, Hefei, 230601, Anhui, China}

\title{Science Done on a Machine by a Machine: AI Agents in Computational Chemistry}

\newcommand{\docstatus}{preprint}\newcommand{\logostrip}{  \raisebox{-0.5\height}{\includegraphics[height=0.85in]{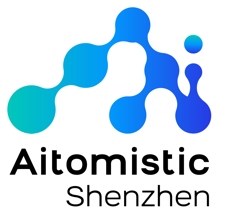}}\hspace{1.6em}  \raisebox{-0.5\height}{\includegraphics[height=0.95in]{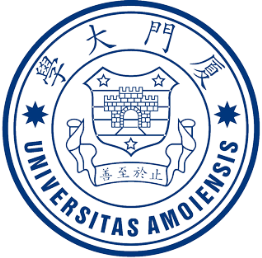}}\hspace{1.6em}  \raisebox{-0.5\height}{\includegraphics[height=0.80in]{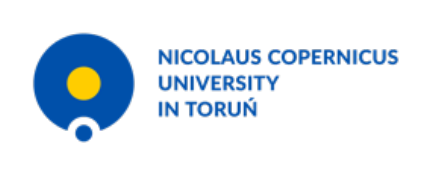}}\hspace{1.6em}  \raisebox{-0.5\height}{\includegraphics[height=0.90in]{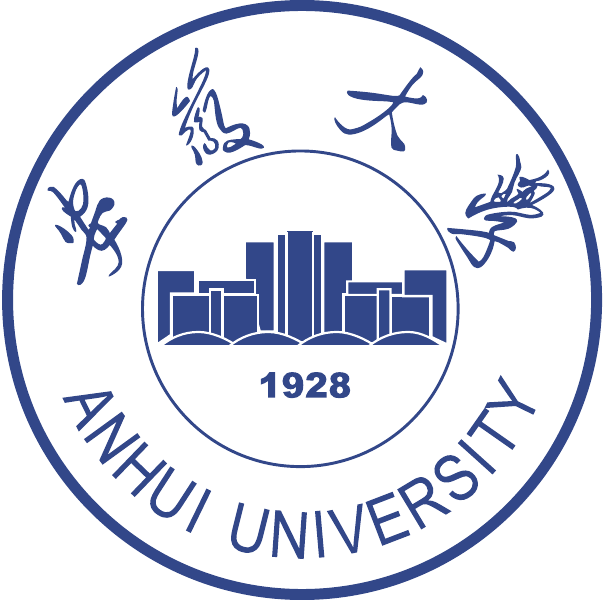}}}
\newcommand{\draftbanner}{}

\newcommand{\Fw}{61}
\newcommand{\RungCalc}{27}
\newcommand{\RungExpt}{29}

\newcommand{\RungUnrecorded}{1}

\newcommand{\PopTwentyFour}{4}
\newcommand{\PopTwentyFive}{17}
\newcommand{\PopTwentySix}{40}
\newcommand{\PopCutoff}{8 August 2026}

\newcommand{\TaskFreeEnergy}{15}

\newcommand{\FreeEnergyStatic}{9}
\newcommand{\FreeEnergySampled}{6}

\newcommand{\NewestDate}{7 August 2026}

\newcommand{\LicStated}{34}
\newcommand{\LicUnclear}{27}
\newcommand{\Hosted}{5}

\newcommand{\RungCalcTwentySixOnly}{13}
\newcommand{\RungExptTwentySixOnly}{23}

\newcommand{\UnlicensedRepos}{7}

\newcommand{\PopFirstDraft}{37}

\begin{document}

\statusblock

\newpage

\begin{abstract}
We are witnessing an explosion of agentic systems for computational chemistry: from four in 2024 to seventeen in 2025 and over sixty now, surveyed here.
What is delegated to these systems is shifting from single calculations to whole \textit{in silico} experiments and even manuscript writing. The ultimate destination is a fully autonomous AI scientist, where the entirety of computational chemistry is performed on a machine by a machine, without human supervision. Our survey shows that these systems are turning into vetted chemistry skills on general-purpose coding agents, and that they must be evaluated not only on their final answers but also on whether their calculations actually support these answers. The role of the human computational chemist is shifting from performing calculations to directing and supervising them, and the field should invest in the judgement that makes the supervision reliable: we propose a reporting standard, an evaluation reproducing published studies, a controlled comparison with general-purpose coding agents, and how to teach.
\end{abstract}

\newpage

\section{What is delegated to a machine}
\label{sec:explosion}

Computational chemistry is one of those research fields that benefited most from the advances in computing hardware and software, which enabled ever-increasing size and time scale, accuracy, and usefulness of simulations.\cite{pople1970approximate,pople1975deficiencies,bartlett2007coupled,kohn1999nobel,behler2007generalized,bartok2010gaussian,smith2017ani,zhang2018deep,batatia2022mace,DeltaLearning,AIQM1,ramakrishnan2014qm9}
This has created a demand for a now sizable number of experts able to perform such simulations, which are highly non-trivial and require years of honing the skills, which in turn creates a substantial barrier to broader adoption of the best practices of computational chemistry.

Nevertheless, many of the single-calculation tasks themselves are rather repetitive and follow standard protocols.
Hence, there was always a desire to automate these repetitive tasks, which was traditionally done with scripts and workflows.\cite{dral2021affordable,wu2025workflowreview,zhang2025chemsmart,migliaro2026chemrefine,MLatom3} Unfortunately, this automation did not diminish the load on human experts, because their work and decision making are then shifted to a higher level: ideation and planning of what and how to calculate, setting up the calculations, writing scripts, monitoring their completion, analyzing the results and failures, adjusting the plan, recalculating failed or erroneous simulations, etc.

\begin{figure}[htbp]
\centering
\includegraphics[width=\textwidth]{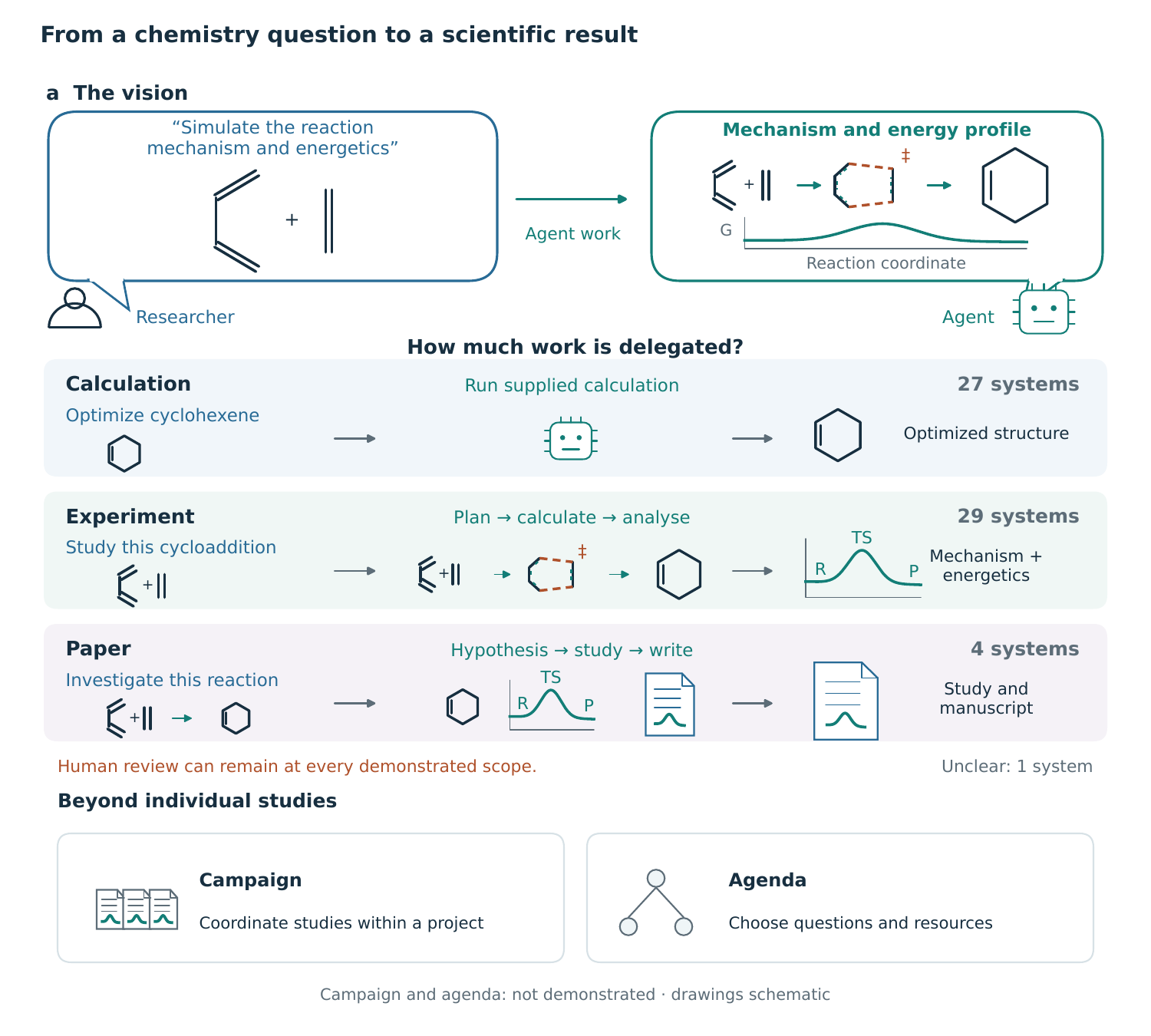}
\caption{From a chemistry question to a scientific result. The opening dialogue illustrates a researcher delegating computational work to an agent. The example is the Diels--Alder cycloaddition of 1,3-butadiene and ethene to cyclohexene. The rows distinguish the largest unit of work delegated: a single calculation; an in silico experiment that includes planning, calculation and analysis; or a study extending from a question to a written manuscript. Counts correspond to the systems surveyed in this Perspective which demonstrate each delegation level. Campaign and agenda are prospective larger scopes that no surveyed system demonstrates; \RungUnrecorded{} system has unclear scope. Chemical structures and energy diagrams are schematic and do not represent numerical results.}
\label{fig:vision}
\end{figure}

One of us (P.O.D.) set out this vision in print~\cite{dral2021affordable} a year before the release of the chat models that started the present wave: machines able to ``perform all these simulations, analyze results and write papers'' and eventually ``make a qualitative jump in our understanding of nature''. It has become possible with the emergence of GPT-4 and modern AI agentic systems, which hold the promise of doing the very tasks that only humans could do: make decisions on the fly (Figure~\ref{fig:vision}), ultimately democratizing computational chemistry for researchers without years of training in computational chemistry.\cite{robinson2026chemworld}

\begin{figure}[!htbp]
\centering
\includegraphics[width=0.93\textwidth]{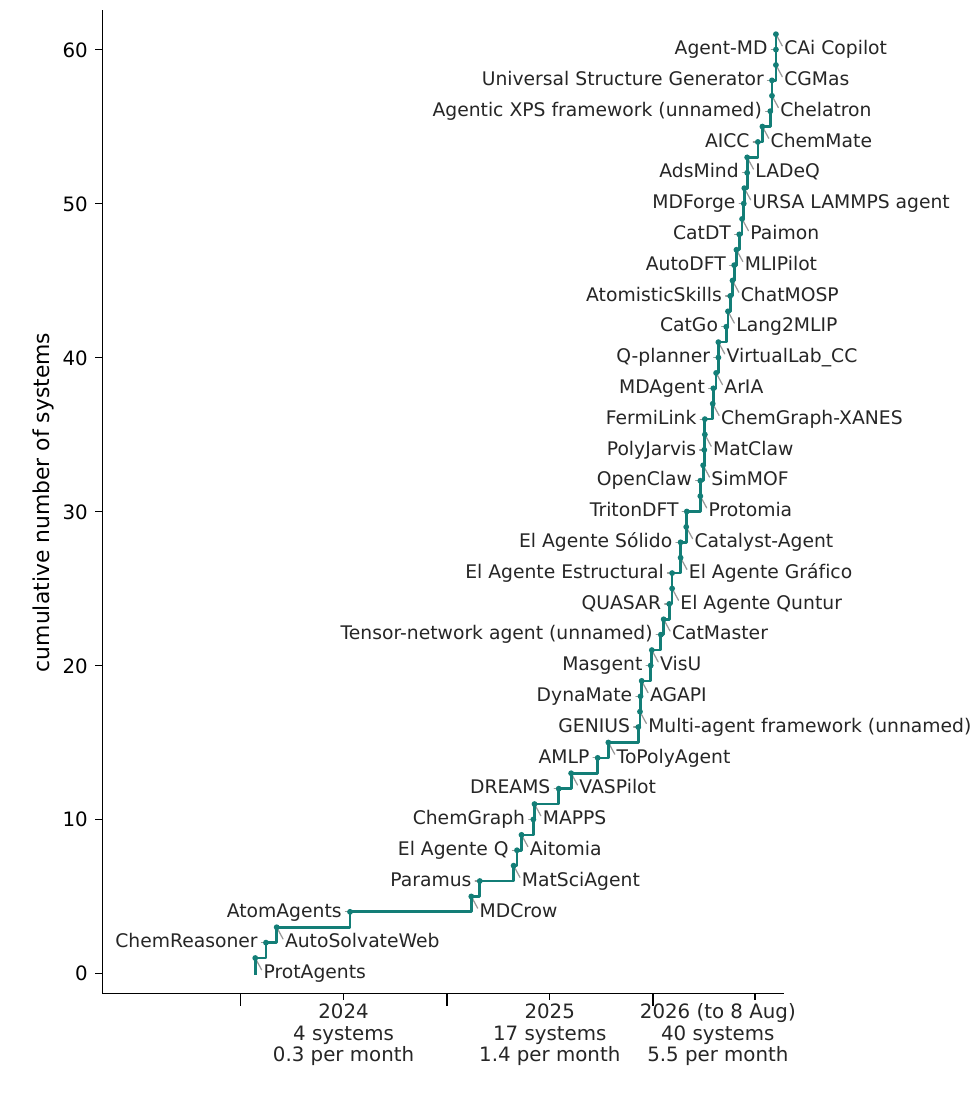}
\caption{Growth of agents for computational chemistry. Each point marks one system at its first public appearance and is named beside the curve. Annual labels give the number of systems and their rate; 2026 runs only to the cutoff of \PopCutoff{}.}
\label{fig:growth}
\end{figure}

Realizing this breakthrough potential of AI, many research groups set out to build agentic systems specialized in computational chemistry, an undertaking that itself became easier with the emergence of the generalist agents we discuss later in this Perspective.
This has led to explosive growth of agents for computational chemistry (Figure~\ref{fig:growth}): \PopTwentyFour{} appear in 2024, \PopTwentyFive{} in 2025, and \PopTwentySix{} up to \PopCutoff{} covered by our survey,
which is scoped to those whose object is an atomistic simulation, set up, run, and interpreted by AI using quantum-chemical, molecular-dynamics or machine-learning-potential calculations on molecules, materials or catalytic surfaces (see Table~\ref{tab:agents} for the full list).

\begin{figure}[!htbp]
\centering
\includegraphics[width=\textwidth]{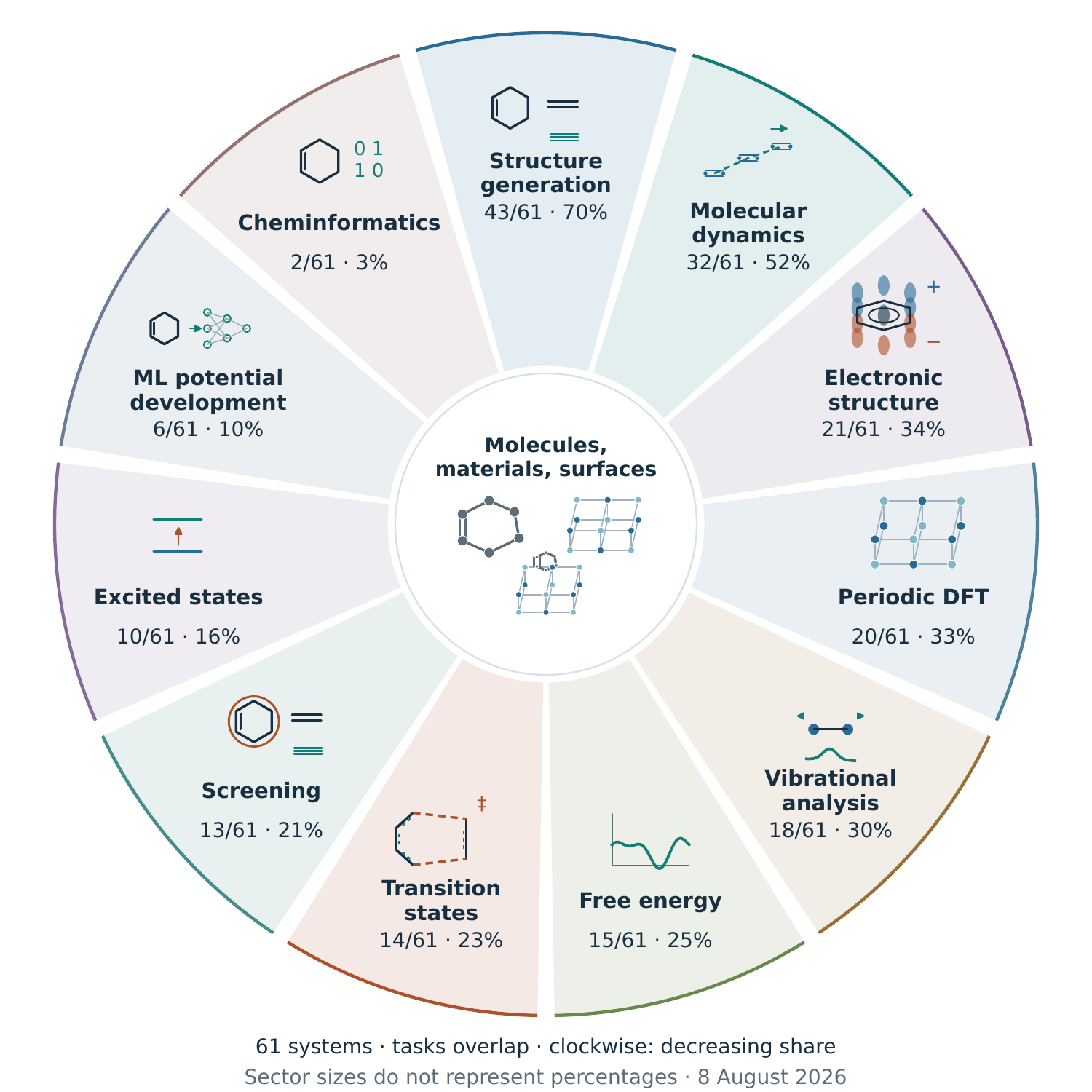}
\caption{Reported computational chemistry capabilities. The tasks are ordered clockwise from the top in descending order of the number of systems reporting each task; sector area is not proportional to the number of systems. A system can report several tasks; percentages use the \Fw{} surveyed systems as the denominator. All chemistry drawings are schematic. Counts use the survey cutoff of \PopCutoff{}.}
\label{fig:coverage}
\end{figure}

The changes in capabilities and architectures are dramatic over such a short period of two and a half years, as we will dissect in this Perspective by performing a meta-analysis of all \Fw{} systems. By now, the reported systems cover a wide range of computational chemistry tasks (Figure~\ref{fig:coverage}).

\begin{sloppypar}
While the early systems were specialized in one or a few tasks, the trend is to generalize them to broader capabilities. The most common task, which is the task zero of any computational workflow, is to generate structures, and it runs from embedding a SMILES string in three dimensions\cite{Aitomia,pham2026chemgraph,chanmungkalakul2026aria} to generating hundreds of hypothetical crystals and screening them.\cite{wang2025dreams,deng2026atomisticskills,ye2026chatmosp} This is followed by molecular dynamics,\cite{campbell2026mdcrow,gadde2025chatbot,chaudhari2026matsciagent,lahouari2025amlp,vriza2025multiagent,guilbert2025dynamate,liu2025masgent,yang2026quasar,zou2026quntur,protomia2026,ding2026openclaw,lee2026simmof,zhao2026polyjarvis,matclaw2026,meng2026fermilink,ma2026mdagent,liu2026catgo,lang2mlip2026,deng2026atomisticskills,yang2026autodft,mlipilot2026,catdt2026,park2026paimon,mdforge2026,somasundaram2026ursa,choi2026cgmas,wang2026agentmd,Zhu2025_VisU,Ding2026_ToPolyAgent,Wang2026_AIComputationalChemist,Liu2026_ChemMateAutonomousResearch} electronic structure,\cite{gadde2025chatbot,zou2025elagente,Aitomia,pham2026chemgraph,zou2026quntur,bai2026grafico,protomia2026,ding2026openclaw,chanmungkalakul2026aria,zhang2026qplanner,kieninger2026virtuallabcc,liu2026catgo,hagai2026ladeq,shen2026xps,summers2026chelatron,Lee2026_AGAPI_Agents,Kumar2026_ElAgenteSolido,Grizzi2026_ChemGraphXANES,Wang2026_AIComputationalChemist,Liu2026_ChemMateAutonomousResearch} and periodic density functional theory\cite{wang2025dreams,liu2025vaspilot,lahouari2025amlp,soleymanibrojeni2025genius,liu2025masgent,chen2026catmaster,yang2026quasar,tritondft2026,protomia2026,lee2026simmof,meng2026fermilink,liu2026catgo,deng2026atomisticskills,yang2026autodft,shen2026xps,Kumar2026_ElAgenteSolido,Grizzi2026_ChemGraphXANES,Wang2026_AIComputationalChemist,Liu2026_ChemMateAutonomousResearch} calculations.
\end{sloppypar}

\begin{sloppypar}
Far fewer systems were designed for screening,\cite{zhu2024chemreasoner,chen2026catmaster,yang2026quasar,catalystagent2026,lee2026simmof,liu2026catgo,deng2026atomisticskills,catdt2026,summers2026chelatron,Lee2026_AGAPI_Agents,Liu2026_ChemMateAutonomousResearch,Wang2026_CaiCopilotReducing} interatomic-potential development,\cite{lahouari2025amlp,chen2026catmaster,matclaw2026,lang2mlip2026,deng2026atomisticskills,mlipilot2026} excited states,\cite{zou2025elagente,Aitomia,li2026tensornetagent,zou2026quntur,bai2026grafico,protomia2026,chanmungkalakul2026aria,zhang2026qplanner,Zhu2025_VisU} and free energies.
While transition states and reaction barriers are the core business of mechanistic computational chemistry, they are represented by only a few systems,\cite{ghafarollahi2024atomagents,paramus2025,Aitomia,chen2026catmaster,zou2026quntur,protomia2026,zhang2026qplanner,kieninger2026virtuallabcc,liu2026catgo,catdt2026,Choi2026_ElAgenteEstructural,Wang2026_AIComputationalChemist,Liu2026_ChemMateAutonomousResearch,Chanmungkalakul2026_UniversalStructureGeneratorPluginAgent} which is not that surprising in hindsight, considering that this is one of the most challenging tasks of computational chemistry.
Free energies appear \TaskFreeEnergy{} times, and how they are obtained separates the survey more sharply than the count does.
The static route accounts for \FreeEnergyStatic{} of them,\cite{zou2025elagente,pham2026chemgraph,chen2026catmaster,zou2026quntur,zhang2026qplanner,kieninger2026virtuallabcc,liu2026catgo,catdt2026,Kumar2026_ElAgenteSolido} converting harmonic frequencies and ideal-gas entropies into a Gibbs energy,\cite{pham2026chemgraph,kieninger2026virtuallabcc} which yields p\textit{K}$_\mathrm{a}$ values,\cite{zou2025elagente,zou2026quntur,zhang2026qplanner} ring strains,\cite{zou2025elagente} deprotonation energies\cite{zou2025elagente,zou2026quntur} and adsorption free-energy diagrams\cite{chen2026catmaster,liu2026catgo,catdt2026} for the price of a frequency calculation.
The other \FreeEnergySampled{} sample the ensemble,\cite{guilbert2025dynamate,ma2026mdagent,deng2026atomisticskills,mdforge2026,Liu2026_ChemMateAutonomousResearch} through umbrella-sampling potentials of mean force,\cite{ma2026mdagent} endpoint MM/PB(GB)SA binding energies,\cite{guilbert2025dynamate} and the alchemical-perturbation and thermodynamic-integration family with Bennett and multistate-Bennett estimators.\cite{mdforge2026}
Taken together, the survey covers structures and dynamics broadly and reaction mechanisms rather thinly.
\end{sloppypar}

\section{How much is delegated to a machine}
\label{sec:landscape}

From the perspective of the evolution of agentic systems, it is important to consider not just \textit{what} they can do, but also \textit{how much} they can accomplish \textit{autonomously}. As a proxy measurement to answer this question via statistical analysis of the systems, we categorize the size of the autonomous chunk of work from a single calculation (e.g., geometry optimization), to an \textit{in silico} experiment (including planning and performing a sequence of calculations and their analysis, etc.), to accomplishing all tasks from ideation up to writing a paper. These demonstrated scopes are illustrated in Figure~\ref{fig:vision}; their counts by year are given in Figure~\ref{fig:temporal}b,c in the Appendix. Beyond that is a bigger campaign scale of autonomy where AI is capable of performing a series of investigations within a scoped project leading to multiple papers and ultimately setting an agenda by itself without any human intervention at all: defining what to investigate, shaping projects, managing resources, and driving them to completion. The latter is when computational chemistry research will be run on a machine by a machine from end to end.

Here, the trend is unmistakable and follows the overall trend in agentic systems: they become more and more capable and autonomous. While the early systems were typically operating on a scale of a single calculation, they quickly shifted to the experiment-scale capabilities, which now constitute the bulk. The current frontier is pushing those systems to the paper level, which is claimed by only a few systems without yet clear evidence that they can produce papers \textit{en masse} without human supervision.
None of the systems reaches the campaign level yet. The closest is ChemMate,\cite{Liu2026_ChemMateAutonomousResearch} which took five separate studies from an idea to a written result, two of them to a manuscript draft, with a human approving the key steps. This is paper-level autonomy done at scale, not yet a research campaign, which would tie such studies into one project leading to several papers; but public reports lag behind what the frontier labs are pursuing at the moment, i.e., such efforts are already actively undertaken, akin to similar undertakings in other fields\cite{lu2024ai,gottweis2026coscientist,boiko2023autonomous}; we can confidently claim this because we pursue such a goal ourselves in our labs. Before the field is solidified at the campaign level, talking about the agenda level is premature, unless we do arrive at a proverbial singularity (superintelligence) faster than we can imagine.

\section{The evolution of architectures enabling the increasing autonomy}
\label{sec:architecture}

\begin{figure}[htbp]
\centering
\includegraphics[width=\textwidth]{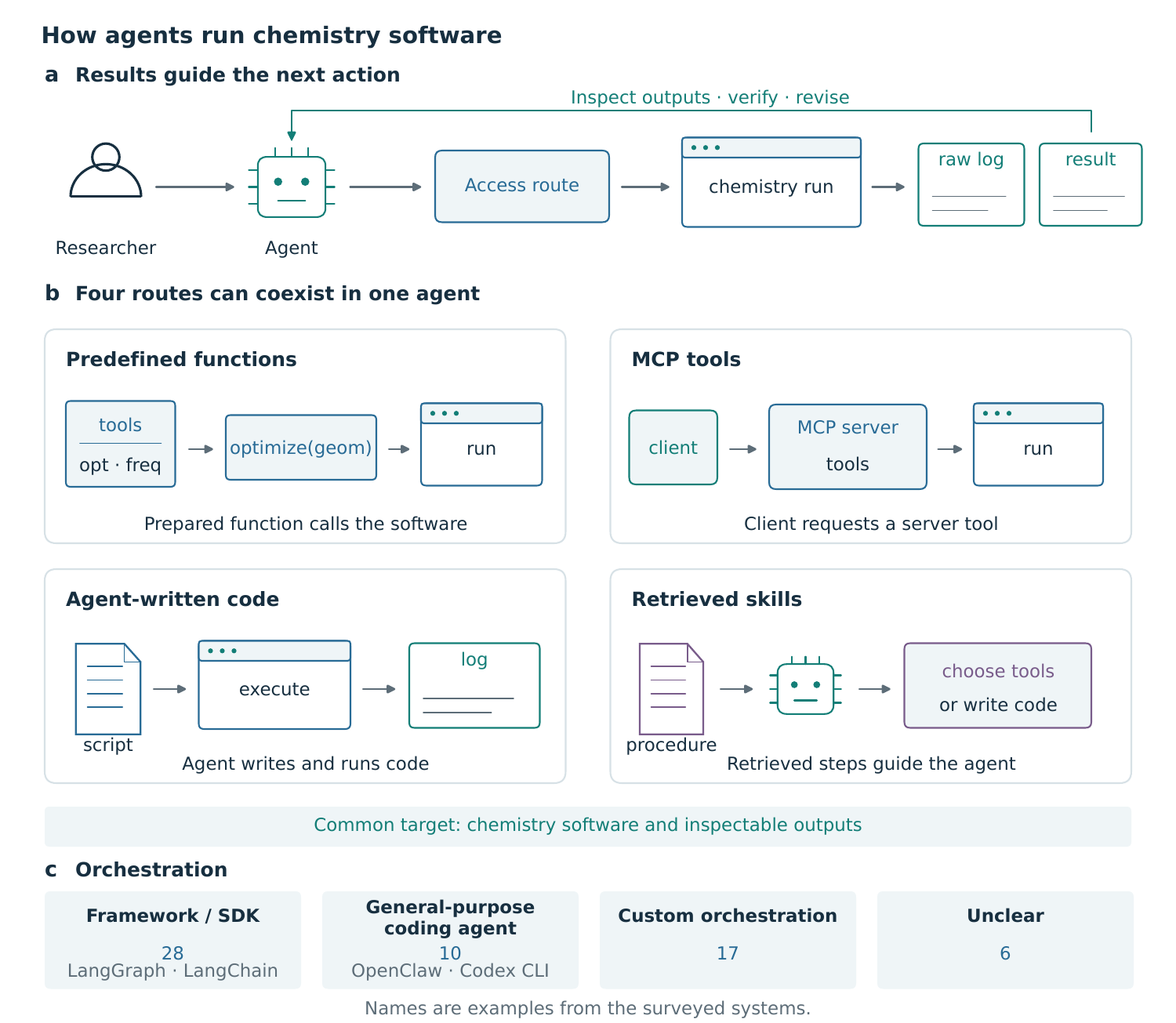}
\caption{How agents run chemistry software. (a) Results inform subsequent actions in an agent loop. (b) Four coexisting routes provide access to chemistry operations: predefined callable functions, tools served through the Model Context Protocol, code written and executed by an agent, and retrieved procedural skills that guide tool use. (c) The orchestration layer is distinct from these routes; counts classify surveyed systems by framework or SDK, general-purpose coding agent, custom loop or unspecified orchestration. Named frameworks and coding agents are examples from the survey, not exhaustive lists. A custom loop names the orchestration implementation, not code written by the agent.}
\label{fig:architecture}
\end{figure}

Pushing agentic systems to higher levels of autonomy depends on the maturity of architectures and the intrinsic capabilities of the underlying AI models (currently mostly large language models, LLMs), with the state-of-the-art models built by private companies, which do not necessarily pursue goals aligned with the requirements for performing autonomous (computational chemistry) research. Hence, our work as experts boils down to repurposing those tools for our research goals, which is complicated by their extremely rapid change, causing a significant lag in development. Our survey clearly reflects this (Figure~\ref{fig:temporal}a): early agentic systems specialized in computational chemistry were based on predefined tool functions (written before the run and shipped with the system),\cite{ghafarollahi2024protagents,zhu2024chemreasoner,ghafarollahi2024atomagents,gadde2025chatbot,campbell2026mdcrow,chaudhari2026matsciagent,zou2025elagente,Aitomia,pham2026chemgraph,wang2025dreams,liu2025vaspilot,lahouari2025amlp,vriza2025multiagent,guilbert2025dynamate,liu2025masgent} which were the only callable tools available to LLMs, whose role was basically to choose the tools and parameters for them, with some departures\cite{soleymanibrojeni2025genius,lahouari2025amlp,yang2026quasar} from this structure when LLMs were fed documentation of the software and asked to generate input files, etc. Later systems increasingly rely on agent-written code (a script the agent composes during the run and executes),\cite{li2026tensornetagent,chen2026catmaster,yang2026quasar,zou2026quntur,bai2026grafico,matclaw2026,meng2026fermilink,ma2026mdagent,mlipilot2026,catdt2026,park2026paimon,mdforge2026,somasundaram2026ursa,hagai2026ladeq} the Model Context Protocol (MCP, letting an agent call tools that were not shipped with it),\cite{liu2025vaspilot,bai2026grafico,catalystagent2026,zhao2026polyjarvis,zhang2026qplanner,liu2026catgo,deng2026atomisticskills} and skills libraries (a set of retrieved written procedures that guide the agent's tool use or code execution) to extend their capabilities.\cite{chen2026catmaster,ding2026openclaw,meng2026fermilink,ma2026mdagent,zhang2026qplanner,kieninger2026virtuallabcc,deng2026atomisticskills,ye2026chatmosp,catdt2026,park2026paimon} Figure~\ref{fig:architecture} distinguishes these routes from the orchestration layer.

The orchestration layer moved accordingly: LangGraph was the popular backbone of the early architectures,\cite{zou2025elagente,Aitomia,pham2026chemgraph,wang2025dreams,zhang2026qplanner,somasundaram2026ursa,shen2026xps,choi2026cgmas,Grizzi2026_ChemGraphXANES,chanmungkalakul2026aria} and now a loose collection of skills sitting on a general-purpose agent takes its place.\cite{ding2026openclaw,meng2026fermilink,deng2026atomisticskills,wang2026agentmd,ma2026mdagent,ye2026chatmosp,Wang2026_AIComputationalChemist,Liu2026_ChemMateAutonomousResearch} Note that skills often do include predefined functions, so the predefined layer is still present in the architecture, and overall, the newer layers accumulate beside the predefined one rather than replacing it.\cite{chen2026catmaster,ding2026openclaw,zhang2026qplanner,catdt2026,park2026paimon} This shift in architecture naturally enabled agentic systems to perform increasingly complex tasks: while initial systems typically could only do a predefined sequence of tasks,\cite{ghafarollahi2024protagents,gadde2025chatbot,ghafarollahi2024atomagents,campbell2026mdcrow,chaudhari2026matsciagent,Aitomia,pham2026chemgraph,liu2025vaspilot,lahouari2025amlp,soleymanibrojeni2025genius,guilbert2025dynamate,liu2025masgent} modern systems typically are capable of working in a loop by letting a result change what runs next.\cite{li2026tensornetagent,chen2026catmaster,yang2026quasar,zou2026quntur,catalystagent2026,lee2026simmof,matclaw2026,meng2026fermilink,lang2mlip2026,deng2026atomisticskills,yang2026autodft,mlipilot2026,catdt2026,park2026paimon,mdforge2026,somasundaram2026ursa,zhang2026adsmind,hagai2026ladeq,summers2026chelatron,wang2026agentmd}

\section{Where do we stand now and what's the end game?}
\label{sec:closing}

At the end of the day, where do we stand now, as of August 18, 2026, when we are writing this? The autonomous tools enabling computational chemistry are here to stay; we do not see going back to hand-writing input files and scripts and manually checking the progress of calculations and analyzing everything by hand. The systems become more capable and trustworthy, as they can work tirelessly and check their own work, while spawning subagents to parallelize work as needed, which is beyond any human capabilities. Despite this, practice and our survey show that agents are currently neither an autonomous scientist nor an obedient calculator: the machine performs the work within boundaries set by the user, and the role of human experts is shifting more towards supervision.

That said, at the level of single calculations the current agentic systems are pretty robust, as our own and other evaluations have shown for some time now.\cite{Aitomia,zou2025elagente,pham2026chemgraph,campbell2026mdcrow} In the benchmark of our first-generation agentic system Aitomia, the success rate was 99.6\% for a low-autonomy single computational task~\cite{Aitomia}. These numbers dropped to 70.9\% and 45.6\% when tasks demanded higher autonomy. Our move from the LangGraph-based Aitomia to Protomia illustrates how rapidly architectures change in our own work. We now use Protomia for experiments and studies extending from ideation to manuscript preparation, although the latter still require substantial human supervision, as illustrated in its documentation\cite{protomia2026docs}. Protomia still awaits systematic evaluation and a formal publication.

Our personal experience with agentic systems, since we started to work on them in 2024, has traced a big arc: while early systems were pretty poor but fascinated us by showing flashes of impressive decision making, now we are frustrated when they do not write a ``perfect'' paper according to our taste, or miss a flaw in their reasoning when planning a new study that is ``obvious'' only to an expert who spent decades honing that judgement. Capabilities of the systems have grown, but our appetites have grown even more. For example, in 2025, we started to use Aitomia for teaching students in our \textit{Computational Chemistry and AI} course and saw that new students by default fall back on it when they need to calculate anything, but its initial versions were not good enough to perform graduate-level research.

This rapid progress of capabilities of our and other agentic systems is to a great extent explained by the improving capabilities of the general-purpose coding agents such as Claude Code, GitHub Copilot, Codex, Qoder, and emerging open-source alternatives. These same tools collapsed the barrier to building agentic systems for computational chemistry and led to their explosion, and they now carry much of what the specialized systems do. This is where the field already stands: chemistry work is increasingly packaged as reusable skills on general-purpose coding agents, with human experts responsible for their validation. Anyone can now simply ask a general-purpose coding agent to get the job done, and an expert who reduces their agentic system to a set of skills and MCP tools makes it callable from any of these agents, which is what we see happening across the field. With a general-purpose coding agent able to do the same job as a specialized system, and autonomous research supported on top of it, as, e.g., in Claude Science, the specialized systems turn into vetted chemistry skills for the general-purpose platforms, and the experts put their effort into these skills and their validation instead of maintaining a standalone application. Both sides benefit: chemistry expertise goes into the skills, and everyone else can use it without years of training in computational chemistry.

Making our agentic systems available online also taught us how strongly adoption depends on habit and trust: people reach for tools that fit the way they already work. Some members of our groups still copy scripts between an online chat and an HPC cluster and run them manually, even when an agent could perform these steps. Many of the other group members prefer to use the general-purpose coding agents to do their research.
This presents an opportunity: those who learn to direct these systems while the transition is still happening will be the ones the field ends up with, and they will reach a scale of materials and questions that was previously out of reach.

Breaking the barrier to performing simulations poses new challenges related to their cost. Paradoxically, delegating the work to a machine does not remove calculation as the bottleneck. Instead, it makes compute more of a bottleneck than ever before, because the cost of atomistic compute is joined by the ``cost of intelligence compute'', which depends on the number and price of the tokens used. An agent can run more calculations than a human would, and it responds faster to failed jobs, which often means rerunning them with different settings, further compounding the need for computing resources. Hence, agentic systems also have to be evaluated in terms of cost, broken down into these two components, and effort has to be put into making agentic systems affordable while keeping them capable. Cost management now becomes one of the essential skills of the human experts driving agentic systems.

To reduce the cost of tokens and ensure privacy, open-weight models deployed locally on a research group's cluster will become as common as installed quantum-chemistry software is today, and they will run the routine work of agentic systems. They do not yet replace the frontier models, which are likely to remain for the steps that need the strongest judgement. In our own group, a local open-weight model already runs around the clock as the dispatcher of routine agentic work, and the frontier models are called only for the critical steps. This keeps the token cost down, in the same way as a cheaper level of theory is chosen to reduce the resources spent on atomistic simulations.

The other obstacle is evaluation: it is becoming genuinely hard to tell where these systems stand. Evaluating a new computational method is possible by running a benchmark on known reference values. Evaluating agentic systems requires human judgement, which is scarce, and the speed of their evolution, with rapidly changing AI models and architectures, makes it practically impossible to compare various systems on an equal footing. Also, any benchmark, once leaked, contaminates the evaluation of the next systems.

Published claims are also increasingly hard to validate, complicated by the cost of such evaluations in tokens and by the fact that many systems disclose neither their code nor an online platform (Table~\ref{tab:agents}).
Whether a reader may legally reuse the code is, for many of these systems, unclear.
\LicStated{} of the \Fw{} systems state a licence, most often MIT.
For the remaining \LicUnclear{} we could find none, and \UnlicensedRepos{} of those are repositories anyone can open today that carry no licence file at all, which leaves a reader code they can read and may not reuse.
Most ship no tests and no continuous integration, and several cannot be run as published at all, through a missing module, a commented-out entry point or a hard-coded path from the author's own machine.
Only \Hosted{} can be tried online without installing anything: TritonDFT,\cite{tritondft2026} VASPilot,\cite{liu2025vaspilot} AutoSolvateWeb,\cite{gadde2025chatbot} AGAPI\cite{Lee2026_AGAPI_Agents} and our own Protomia,\cite{protomia2026} which is a hosted service and not an open-source release.

The end game that the field pursues is agentic systems that run unattended around the clock for weeks, months and years, carrying a research campaign and then an agenda. None of the surveyed systems has achieved this yet. Such agentic systems may evolve themselves as they go, improving their own skills, tools and procedures from what their runs teach them. Human supervision will gradually shift to the level of the project rather than of the calculation. This destination needs two of the recommendations that follow: an evaluation that checks whether the calculations support the answer, because nobody will read every result of a run that lasts for months, and a reported cost, which limits how long an unattended run can be afforded.

\section{What the field should do}
\label{sec:recommendations}

\begin{figure}[htbp]
\centering
\includegraphics[width=\textwidth]{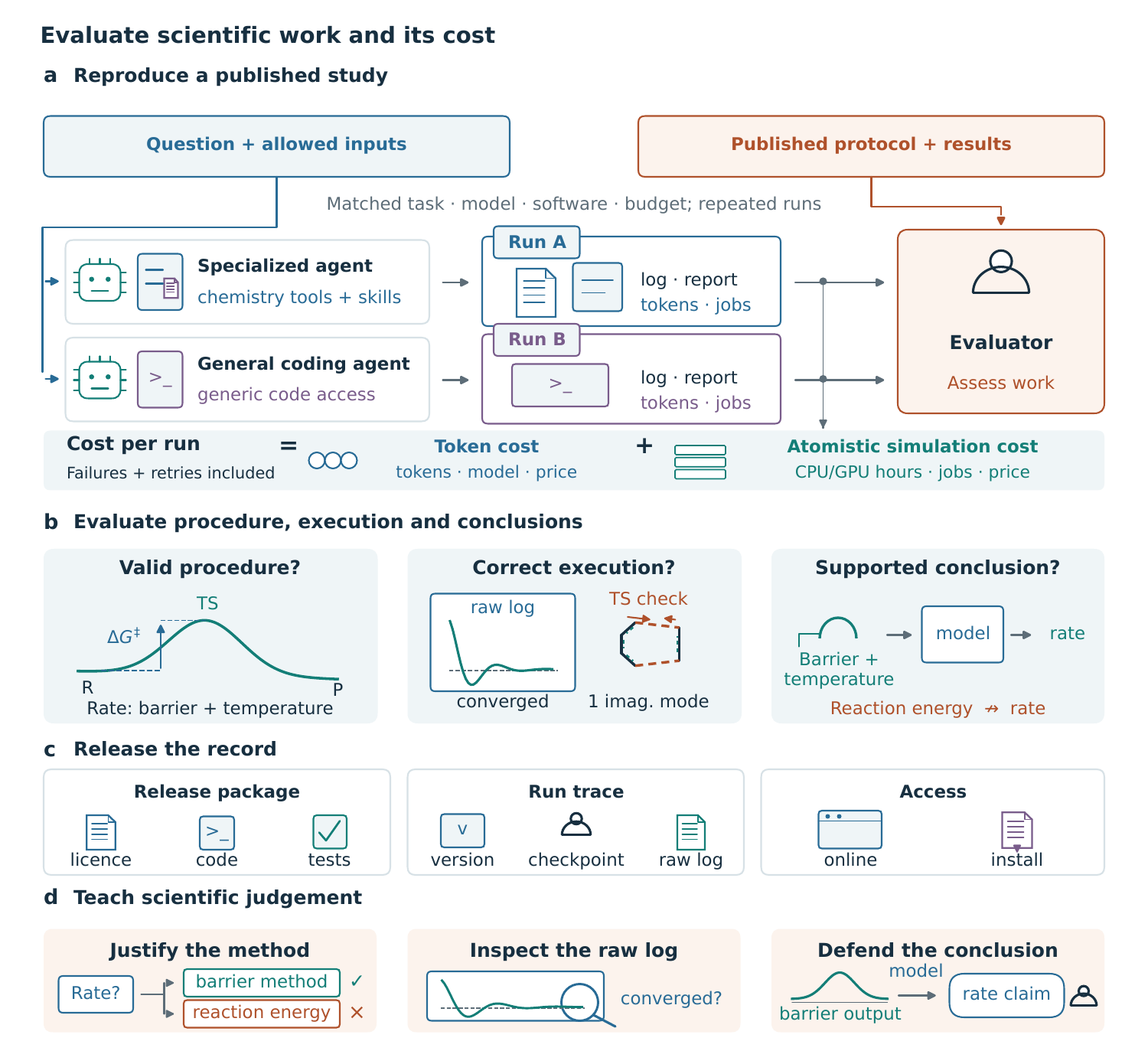}
\caption{Evaluation of agentic scientific work and its cost. (a) The proposed evaluation asks agents to reproduce published computational studies from the question to the report. Agents receive the question and defined allowed inputs; evaluators use the published protocol and results as reference evidence. A specialized system and a general-purpose coding agent without chemistry skills receive matched tasks, base models, software access and budgets, with repeated runs. The cost of each run is token cost plus atomistic simulation cost, including failed attempts and retries. (b) Evaluation should check whether the procedure is valid, the computation is executed correctly, and the conclusions drawn are correct. (c) Release records should include licence, runnable code and tests, model version, run traces and access; reports should also disclose scope and human interventions. (d) Teaching should develop the corresponding abilities to justify method choices, inspect raw outputs, and defend conclusions.}
\label{fig:evaluation}
\end{figure}

Here we outline our recommendations and our view on how agentic systems should be reported, how they should be evaluated, and how students should be taught to use them. These recommendations are summarized in Figure~\ref{fig:evaluation}.

We strongly recommend that works reporting an agentic system clearly disclose the licence and ship tests with code that runs as published, to ensure reproducibility. It would also be helpful to state how much can be delegated to a system (a single calculation, an \textit{in silico} experiment, a paper, a campaign or an agenda), whether it can be used online or needs installation (and what is required for it). Affordability is another important consideration, and hence the papers should report the token cost and the atomistic simulation cost of every run, including failed attempts and retries. Our survey shows how far the field is from this: most systems ship no tests, several cannot be run as published, and most report neither cost. Stating these items in every paper will make agentic systems more accessible and any two of them comparable.

Evaluation of agentic systems is another important issue, and we suggest grading them on three separable axes by checking whether the agent: 1) chose a valid procedure, 2) executed the computation correctly, and 3) drew conclusions that follow from what was actually computed. One of the standard evaluations in the future should consist of validating how well the agents can reproduce the published computational chemistry studies that humans did before agents existed, from the question through the calculations to the written report, with the published Supporting Information as the reference. Such reproductions have already started in computational chemistry: FermiLink reproduced figures of published simulation studies across nine research domains,\cite{meng2026fermilink} CatGo reproduced a published study of a nickel catalyst,\cite{liu2026catgo} and AICC answered the referees' computational requests on five published \textit{Nature Communications} papers, whose original authors' calculations served as the human reference: one of its two LLM evaluators scored the agent's reports above the authors' own calculations and the other below them.\cite{Wang2026_AIComputationalChemist} The AutoMat benchmark poses the same task for claims from computational materials science.\cite{huang2026automat} Outside chemistry, such reproduction benchmarks already exist for machine-learning research\cite{starace2025paperbench} and for re-running the code of published studies.\cite{siegel2024corebench} That the published answer is in the training data of the model does not matter, because the calculation is what is graded, and this also alleviates the issue with the contamination of leaked benchmarks. Each task should be run repeatedly, because the same task can succeed on one run and fail on another due to the non-determinism of AI models,\cite{Aitomia} and the evaluation should report the token and atomistic simulation cost of every run. Such an evaluation grades the whole chain from question to report that the paper rung requires, and it can be repeated on every new model and agentic system as they appear, which will help to build robust agentic systems capable of making expert calculations accessible to a wider chemistry community.

The same task set should also be used for a controlled comparison of any specialized agentic system with a general-purpose coding agent that is given no chemistry skills at all. This comparison shows whether specialization adds anything to what general-purpose coding agents can already do and whether it is worth the effort. Specialization may even, paradoxically, hurt the agent by biasing it towards specific workflows that are not optimal for the task at hand. We also emphasize the importance of benchmarking various backend AI models, preferably also local open-weight ones (for reproducibility and cost reduction), to gain insight into how much the quality of the agent depends on the backend and which one is recommended.

Agentic systems have a big impact on training the next generation of scientists, and how we adjust our teaching is of paramount importance. We should train the students to justify computational choices, inspect computational evidence, and trace conclusions to calculation outputs, and evaluations should test these abilities alongside task completion and final-answer accuracy. In our \textit{Computational Chemistry and AI} course, new students fall back on the agentic system by default whenever they need to calculate anything, so what they must learn is to direct it and to check what it returns. Students trained this way become the supervisors that the delegation of whole studies requires.

The same recommendations apply beyond chemistry. The delegation scale, the three grading axes and the convergence onto general-purpose coding agents hold for any simulation science and for autonomous-science agents in general.\cite{lu2024ai,gottweis2026coscientist,boiko2023autonomous}

The question for the field is no longer whether agentic systems will become standard tools in computational chemistry, but what kind of scientists they will make of us. On the evidence of our own survey and our own lab, the answer looks like this: a junior collaborator that the expert supervises, a tool that multiplies what a student can reach, and a partner that frees the expert for the judgement that the machine cannot yet supply. This Perspective was itself prepared by agentic systems, one of them (Protomia) surveyed here, under the authors' supervision (see Methods). AI is here to help us do more and better science, and the researchers who learn to direct it well will make more discoveries, not fewer.

\section*{Author contributions}
A.U. conceived the Perspective and drafted its first version. H.N. extended and revised that draft and prepared the initial survey of the agentic systems. P.O.D. set the argument and scope, built the survey database in its present form, and wrote the final manuscript with assistance from the AI Dral Group.
The AI Dral Group, the group's collective of specialized AI research agents, assisted with the survey database, intermediate drafts, figures and final polishing under P.O.D.'s direction and the authors' verification. All authors discussed, revised and approved the manuscript.

\section*{Conflict of interest}
Two of the systems surveyed here, Aitomia and Protomia, are our own, and Protomia was used in
preparing this article, as described in Methods. They were scored on the same rules as every
other entry, applied from the same public sources, and those rules are stated in Methods. P.O.D.\ is a co-founder of and holds equity in Aitomistic, which develops
Protomia, and is an author of Aitomia; the remaining authors declare no competing financial
interest.

\section*{Data availability}
No data were generated in this work, and the records behind the survey are not deposited,
because they consist largely of text quoted verbatim from the surveyed papers, which is
copyrighted by its publishers; the survey itself is given in full in Table~\ref{tab:agents}.

\section*{Methods}

We surveyed the literature to a cutoff of \PopCutoff{}, working outward from the reference lists
of the three existing surveys and from arXiv and ChemRxiv listings in the relevant subject
classes, and adding systems that colleagues brought to our attention, so the corpus is not the
output of a single reproducible query. Systems are ordered by first public
appearance, preprint or journal, whichever came first. Frameworks appeared while this article was
being prepared, so Table~\ref{tab:agents} is a snapshot and \PopTwentySix{} is a floor rather
than a year's total, its most recent member having appeared on \NewestDate{}.

Two standards are used, and they are stated separately rather than blended. A scientific task
counts where the authors report their own system performing it, in their own tense: an
implemented capability counts even where the worked example is missing, whereas a task placed in
future work, or one performed by a program they merely cite, does not. Delegation is held to the
stricter standard and counts only what a paper demonstrates in its body, because the claim made
from it is about how far the field has actually got.

The rule selects systems, not tasks: a system in the survey for the atomistic-simulation workflow it automates is then
recorded for every task it performs, cheminformatics included. Agents for adjacent work are cited
where they bear on the argument and deliberately not tabulated --- retrosynthesis, property
prediction with no underlying simulation, robotic synthesis, materials informatics in which
first-principles results appear only as an external check, and systems built to do science in
general rather than chemistry in
particular\cite{Legrand2026_MimosaFrameworkEvolving} --- because including them would broaden the
comparison until it said nothing specific about computational chemistry.

Generative AI was used substantively throughout. The survey was assembled, the database
built and the intermediate versions of this article written by the AI Dral Group, an AI
research-agent collective independent of the underlying runtime; here it operated on
Claude Code, Protomia\cite{protomia2026}, and Codex for literature retrieval, citation verification and date
resolution, Protomia being itself one of the systems surveyed in Table~\ref{tab:agents}.
P.O.D.\ wrote the final version with assistance from the collective.
Figure~\ref{fig:growth} is drawn from the database on every build. Figures~\ref{fig:vision}, \ref{fig:coverage} and \ref{fig:architecture} contain database-derived counts; Figures~\ref{fig:vision}, \ref{fig:architecture} and \ref{fig:evaluation} also contain schematic illustrations. The source code for the five figures is re-run on each build; Figure~\ref{fig:temporal} in the Appendix retains the database-driven temporal survey panels. The final figures are prepared by Codex (and some initial versions by Claude Code) under the authors' direction. The diagrams illustrate chemistry and procedures, not computed results. The other scripts, and
the analysis and gate code behind every count and table, were written by the same agents
under the authors' direction: what they produce is fixed by the data rather than by a model,
whereas the prose is not. The authors directed and verified every step, checking the
rendered figures and the typeset page rather than the source, and take full responsibility
for the content.

Each tabulated field is recorded against a verbatim passage and a location in the system's own
source, read in full rather than from secondary descriptions, except where the fact is an absence,
which is noted in the quote slot; our own Protomia, which has no paper yet, is evidenced from its
documentation and from its use in our groups and by other users. Every reference was checked against its primary record (publisher front matter,
Crossref, or the arXiv API), a pass that corrected author lists, an identifier pointing to an
unrelated paper, and characterizations that did not match the cited work's findings. Pre-2015
books, reports and printed proceedings carry no persistent identifier and are cited from the
printed record alone.

Three limits could each move a number here. We read papers rather than running systems, so every
count measures what a paper shows and a system that works better than it writes is undercounted.
We cannot estimate our own recall: the survey grew from \PopFirstDraft{} to \Fw{} systems during
preparation, mostly from papers that already existed and our search had missed, so it is a lower
bound.
Across all years,
\textit{in silico} experiments outnumber single calculations by only two systems (\RungExpt{} to \RungCalc{}),
so a different judgement on a few multi-step workflows could reverse that order; among the systems
released in 2026, experiments lead clearly (\RungExptTwentySixOnly{} to \RungCalcTwentySixOnly{})
whichever way those workflows are placed.

\appendix
\renewcommand{\thefigure}{A\arabic{figure}}
\setcounter{figure}{0}

\begin{landscape}
\begin{figure}[p]
\section*{Appendix: the surveyed systems}
\label{page:appendix}
\centering
\includegraphics[width=0.88\linewidth]{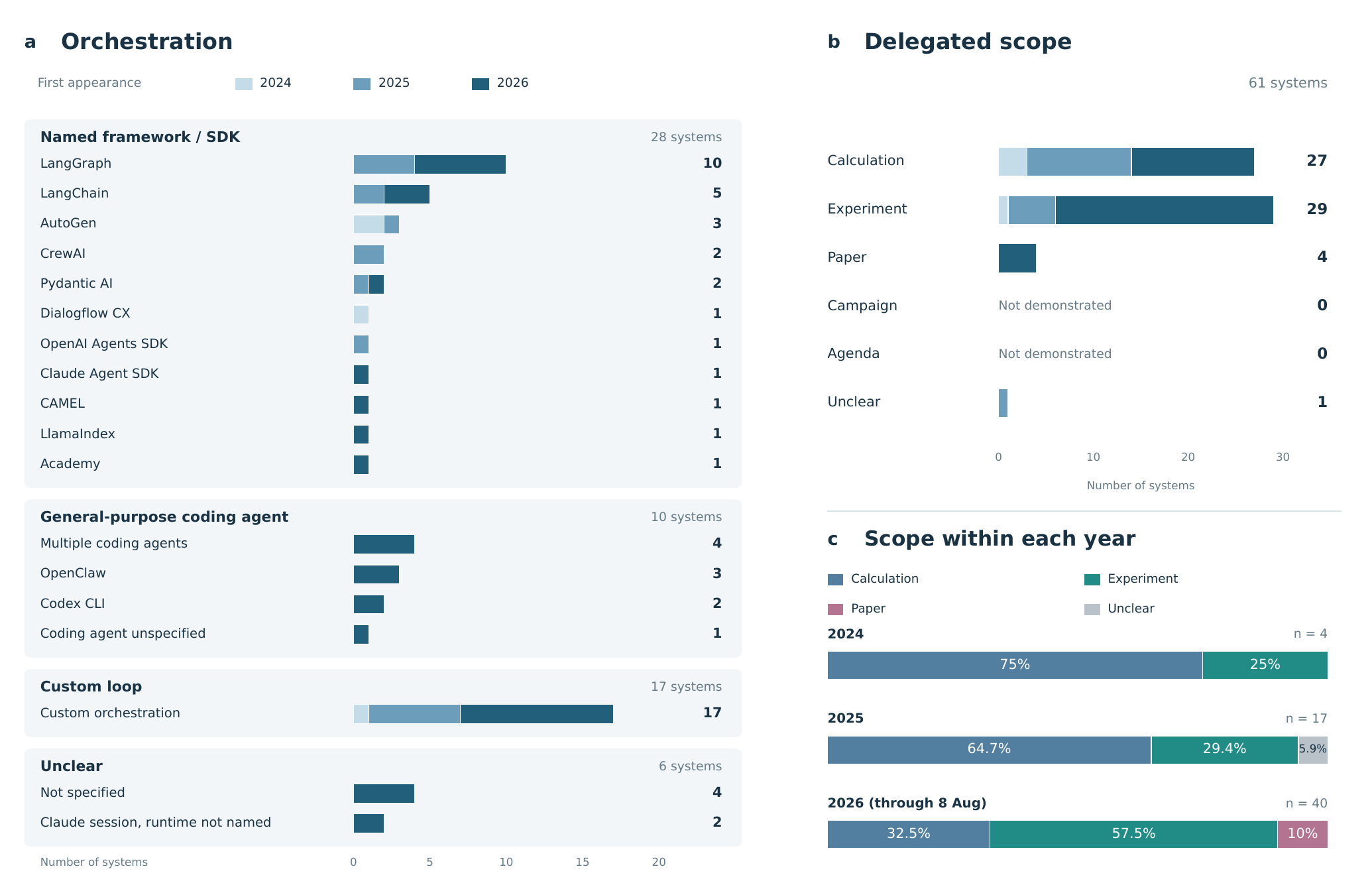}
\caption{Architectures and delegation over time. (a) Orchestration of the surveyed systems, grouped by year of first appearance. (b) Number of systems at each delegation level, grouped by year. (c) Percentage of systems at each delegation level within each year. The survey cutoff is \PopCutoff{}.}
\label{fig:temporal}
\end{figure}
\end{landscape}

Table~\ref{tab:agents} lists every system in the survey with the evidence each column is read from. It is placed here rather than in the body because of its length.

\begingroup
\scriptsize
\setlength{\tabcolsep}{4pt}\begin{longtable}{@{}c >{\raggedright\arraybackslash}p{2.3cm} l >{\raggedright\arraybackslash}p{2.7cm} >{\raggedright\arraybackslash}p{2.9cm} >{\raggedright\arraybackslash}p{1.6cm} l@{}}
\caption{The \Fw{} agentic systems for atomistic simulation surveyed here, ordered by date of first public appearance and surveyed to a cutoff of \PopCutoff{}. Scientific tasks are those the system performs on its authors' own account, read paper by paper: an implemented capability is recorded even where the worked example is missing, but a task the authors place in future work, or one performed by a program they merely cite, is not. The Delegation column is what the paper demonstrates in its body; aspirations and future work are deliberately not tabulated. That column uses the five-position delegation scale defined above; a system that occupies more than one position at once carries a compound entry and is counted at the larger. Architecture names how the agent reaches its capabilities and what runs it: predefined tool functions, the Model Context Protocol, retrieved skills or agent-written code, then, after the separator, the software framework the agent itself runs on. The tool tokens are predefined functions, the Model Context Protocol, retrieved procedural skills, agent-written code, or a combination. A dash marks a system whose agents have no callable tool functions. A question mark marks a value the paper does not record, and n/a a value for which no source exists, the system having no paper at all.}
\label{tab:agents}\\
\toprule
\textbf{\#} & \textbf{Framework} & \textbf{Date} & \textbf{Scientific tasks} & \textbf{Architecture} & \textbf{Delegation} & \textbf{Ref.} \\
\midrule
\endfirsthead
\multicolumn{7}{c}{\footnotesize\tablename~\thetable{} (continued)}\\
\toprule
\textbf{\#} & \textbf{Framework} & \textbf{Date} & \textbf{Scientific tasks} & \textbf{Architecture} & \textbf{Delegation} & \textbf{Ref.} \\
\midrule
\endhead
\midrule
\multicolumn{6}{r}{\footnotesize\itshape continued on next page}\\
\endfoot
\bottomrule
\endlastfoot
1 & \texttt{ProtAgents} & 2024-01-27 & struct.\ gen., vib. & fixed $\cdot$ AutoGen & calculation & \cite{ghafarollahi2024protagents} \\
2 & \texttt{ChemReasoner} & 2024-02-15 & screening, struct.\ gen. & fixed $\cdot$ own harness & calculation & \cite{zhu2024chemreasoner} \\
3 & \texttt{AutoSolvateWeb} & 2024-03-05 & struct.\ gen., MD, elec.\ struct. & fixed $\cdot$ Dialogflow & calculation & \cite{gadde2025chatbot} \\
4 & \texttt{AtomAgents} & 2024-07-13 & TS, struct.\ gen. & fixed $\cdot$ AutoGen & experiment & \cite{ghafarollahi2024atomagents} \\
5 & \texttt{MDCrow} & 2025-02-13 & MD & fixed $\cdot$ LangChain & calculation & \cite{campbell2026mdcrow} \\
6 & \texttt{Paramus} & 2025-02-28 & elec.\ struct., vib., TS, excited, MD, periodic DFT, free energy, struct.\ gen., screening, cheminf. & fixed, MCP, skills $\cdot$ own harness & n/a & \cite{paramus2025} \\
7 & \texttt{MatSciAgent} & 2025-04-29 & struct.\ gen., MD & fixed $\cdot$ LangChain & calculation & \cite{chaudhari2026matsciagent} \\
8 & \texttt{El Agente Q} & 2025-05-05 & elec.\ struct., excited, free energy, struct.\ gen., vib. & fixed $\cdot$ LangGraph & experiment & \cite{zou2025elagente} \\
9 & \texttt{Aitomia} & 2025-05-13 & elec.\ struct., excited, struct.\ gen., TS, vib. & fixed $\cdot$ LangGraph & experiment & \cite{Aitomia} \\
10 & \texttt{ChemGraph} & 2025-06-03 & elec.\ struct., free energy, struct.\ gen., vib. & fixed $\cdot$ LangGraph & calculation & \cite{pham2026chemgraph} \\
11 & MAPPS & 2025-06-05 & struct.\ gen. & generated $\cdot$ own harness & experiment & \cite{Zhou2025_TowardGreaterAutonomy} \\
12 & \texttt{DREAMS} & 2025-07-18 & periodic DFT, struct.\ gen. & fixed $\cdot$ LangGraph & experiment & \cite{wang2025dreams} \\
13 & \texttt{VASPilot} & 2025-08-09 & periodic DFT & fixed, MCP $\cdot$ CrewAI & calculation & \cite{liu2025vaspilot} \\
14 & \texttt{AMLP} & 2025-09-25 & MLIP dev., periodic DFT, MD & fixed, -- $\cdot$ own harness & calculation & \cite{lahouari2025amlp} \\
15 & ToPolyAgent & 2025-10-14 & MD, struct.\ gen. & fixed $\cdot$ CrewAI & calculation & \cite{Ding2026_ToPolyAgent} \\
16 & \texttt{GENIUS} & 2025-12-06 & periodic DFT & -- $\cdot$ own harness & calculation & \cite{soleymanibrojeni2025genius} \\
17 & Multi-agent framework (unnamed) & 2025-12-09 & MD, vib., struct.\ gen. & fixed $\cdot$ AutoGen & calculation & \cite{vriza2025multiagent} \\
18 & \texttt{DynaMate} & 2025-12-10 & MD, free energy & fixed $\cdot$ own harness & calculation & \cite{guilbert2025dynamate} \\
19 & AGAPI & 2025-12-12 & elec.\ struct., struct.\ gen., screening & fixed $\cdot$ OpenAI SDK & experiment & \cite{Lee2026_AGAPI_Agents} \\
20 & \texttt{Masgent} & 2025-12-28 & MD, periodic DFT, struct.\ gen. & fixed $\cdot$ pydantic-ai & calculation & \cite{liu2025masgent} \\
21 & VisU & 2025-12-30 & MD, excited & fixed, generated $\cdot$ own harness & calculation & \cite{Zhu2025_VisU} \\
22 & Tensor-network agent (unnamed) & 2026-01-15 & excited & fixed, generated $\cdot$ LangChain & experiment & \cite{li2026tensornetagent} \\
23 & \texttt{CatMaster} & 2026-01-20 & free energy, MLIP dev., periodic DFT, screening, struct.\ gen., TS, vib. & fixed, skills, generated $\cdot$ LangChain & paper & \cite{chen2026catmaster} \\
24 & \texttt{QUASAR} & 2026-01-30 & periodic DFT, MD, screening & generated, -- $\cdot$ LangChain & experiment & \cite{yang2026quasar} \\
25 & \texttt{El Agente Quntur} & 2026-02-04 & elec.\ struct., excited, free energy, MD, struct.\ gen., TS, vib. & fixed, generated $\cdot$ own harness & experiment & \cite{zou2026quntur} \\
26 & El Agente Estructural & 2026-02-04 & struct.\ gen., TS & fixed, generated $\cdot$ ? & calculation & \cite{Choi2026_ElAgenteEstructural} \\
27 & \texttt{El Agente Gr\'afico} & 2026-02-19 & elec.\ struct., vib., excited, struct.\ gen. & fixed, MCP, generated $\cdot$ pydantic-ai & calculation & \cite{bai2026grafico} \\
28 & El Agente Sólido & 2026-02-19 & elec.\ struct., vib., periodic DFT, free energy, struct.\ gen. & fixed $\cdot$ own harness & calculation & \cite{Kumar2026_ElAgenteSolido} \\
29 & \texttt{Catalyst-Agent} & 2026-03-01 & screening, struct.\ gen. & MCP $\cdot$ Codex CLI & experiment & \cite{catalystagent2026} \\
30 & \texttt{TritonDFT} & 2026-03-02 & periodic DFT, vib. & fixed $\cdot$ own harness & calculation & \cite{tritondft2026} \\
31 & \texttt{Protomia} & 2026-03-26 & MD, elec.\ struct., excited, periodic DFT, TS, vib. & fixed $\cdot$ n/a & experiment, paper & \cite{protomia2026} \\
32 & \texttt{OpenClaw} & 2026-03-26 & struct.\ gen., elec.\ struct., MD & fixed, skills $\cdot$ OpenClaw & calculation & \cite{ding2026openclaw} \\
33 & \texttt{SimMOF} & 2026-03-31 & periodic DFT, MD, screening, struct.\ gen. & fixed $\cdot$ own harness & experiment & \cite{lee2026simmof} \\
34 & \texttt{PolyJarvis} & 2026-04-02 & struct.\ gen., MD & MCP $\cdot$ Claude, ? & calculation & \cite{zhao2026polyjarvis} \\
35 & \texttt{MatClaw} & 2026-04-03 & MLIP dev., MD & fixed, generated $\cdot$ own harness & experiment & \cite{matclaw2026} \\
36 & \texttt{FermiLink} & 2026-04-03 & MD, periodic DFT & skills, generated $\cdot$ any coding agent & experiment & \cite{meng2026fermilink} \\
37 & ChemGraph-XANES & 2026-04-17 & elec.\ struct., periodic DFT & fixed $\cdot$ LangGraph & calculation & \cite{Grizzi2026_ChemGraphXANES} \\
38 & \texttt{MDAgent} & 2026-04-18 & MD, free energy & skills, generated $\cdot$ OpenClaw & experiment & \cite{ma2026mdagent} \\
39 & ArIA & 2026-04-23 & elec.\ struct., excited, struct.\ gen. & fixed $\cdot$ LangGraph & calculation & \cite{chanmungkalakul2026aria} \\
40 & \texttt{Q-planner} & 2026-04-27 & elec.\ struct., excited, free energy, struct.\ gen., TS, vib. & fixed, MCP, skills $\cdot$ LangGraph & calculation & \cite{zhang2026qplanner} \\
41 & \texttt{VirtualLab\_CC} & 2026-04-27 & elec.\ struct., free energy, struct.\ gen., TS, vib. & skills $\cdot$ Claude, ? & experiment & \cite{kieninger2026virtuallabcc} \\
42 & CatGo & 2026-05-11 & MD, elec.\ struct., free energy, periodic DFT, screening, struct.\ gen., TS, vib. & MCP $\cdot$ any coding agent & experiment & \cite{liu2026catgo} \\
43 & \texttt{Lang2MLIP} & 2026-05-14 & MLIP dev., struct.\ gen., MD & ? $\cdot$ Claude SDK & experiment & \cite{lang2mlip2026} \\
44 & \texttt{AtomisticSkills} & 2026-05-18 & struct.\ gen., screening, MD, periodic DFT, MLIP dev., free energy & MCP, skills $\cdot$ any coding agent & experiment & \cite{deng2026atomisticskills} \\
45 & ChatMOSP & 2026-05-22 & struct.\ gen. & skills $\cdot$ OpenClaw & experiment & \cite{ye2026chatmosp} \\
46 & \texttt{AutoDFT} & 2026-05-25 & MD, periodic DFT, vib. & fixed $\cdot$ own harness & experiment & \cite{yang2026autodft} \\
47 & \texttt{MLIPilot} & 2026-05-29 & MLIP dev., MD & fixed, generated $\cdot$ own harness & experiment & \cite{mlipilot2026} \\
48 & \texttt{CatDT} & 2026-06-03 & free energy, MD, screening, struct.\ gen., TS & fixed, skills, generated $\cdot$ CAMEL & experiment & \cite{catdt2026} \\
49 & \texttt{Paimon} & 2026-06-08 & MD, struct.\ gen. & fixed, skills, generated $\cdot$ LlamaIndex & experiment & \cite{park2026paimon} \\
50 & \texttt{MDForge} & 2026-06-11 & free energy, MD & generated $\cdot$ own harness & paper & \cite{mdforge2026} \\
51 & \texttt{URSA} LAMMPS agent & 2026-06-12 & MD, struct.\ gen. & fixed, generated $\cdot$ LangGraph & experiment & \cite{somasundaram2026ursa} \\
52 & \texttt{AdsMind} & 2026-06-17 & struct.\ gen. & fixed $\cdot$ ? & calculation & \cite{zhang2026adsmind} \\
53 & \texttt{LADeQ} & 2026-06-17 & elec.\ struct. & generated $\cdot$ own harness & experiment & \cite{hagai2026ladeq} \\
54 & AICC & 2026-07-06 & elec.\ struct., periodic DFT, MD, TS, vib., struct.\ gen. & fixed $\cdot$ any coding agent & experiment & \cite{Wang2026_AIComputationalChemist} \\
55 & ChemMate & 2026-07-14 & elec.\ struct., TS, periodic DFT, MD, free energy, screening, cheminf., vib., struct.\ gen. & fixed, skills, generated $\cdot$ coding agent & paper & \cite{Liu2026_ChemMateAutonomousResearch} \\
56 & Agentic XPS framework (unnamed) & 2026-07-28 & elec.\ struct., periodic DFT, struct.\ gen. & fixed $\cdot$ LangGraph & calculation & \cite{shen2026xps} \\
57 & \texttt{Chelatron} & 2026-07-31 & struct.\ gen., screening, elec.\ struct. & fixed $\cdot$ Academy & experiment & \cite{summers2026chelatron} \\
58 & Universal Structure Generator & 2026-07-31 & struct.\ gen., TS & fixed $\cdot$ ? & calculation & \cite{Chanmungkalakul2026_UniversalStructureGeneratorPluginAgent} \\
59 & \texttt{CGMas} & 2026-08-07 & MD, struct.\ gen. & fixed $\cdot$ LangGraph & calculation & \cite{choi2026cgmas} \\
60 & \texttt{Agent-MD} & 2026-08-07 & MD, struct.\ gen. & fixed $\cdot$ Codex CLI & experiment & \cite{wang2026agentmd} \\
61 & CAi Copilot & 2026-08-07 & struct.\ gen., screening & fixed, skills $\cdot$ own harness & experiment & \cite{Wang2026_CaiCopilotReducing}
\\
\end{longtable}
\endgroup

\begin{acknowledgement}
The authors thank Jinming Hu for discussions about agentic systems.
Generative AI was used substantively in this work; how, and with what verification, is set out in Methods.
A.U. acknowledges funding from the National Natural Science Foundation of China (No. W2433037) and the Natural Science Foundation of Anhui Province (No. 2408085QA002). P.O.D. acknowledges funding by the projects for International Senior Scientists (Project No.: W2531013) and for Outstanding Youth Scholars (Overseas, 2021) of the National Natural Science Foundation of China, via the Lab project of the State Key Laboratory of Physical Chemistry of Solid Surfaces.

This work was supported by the PIAST-AI Factory project -- PIAST AIF, co-financed by the EuroHPC Joint Undertaking through the European Union's Digital Europe Programme, and by Poland as the Participating Country, under Grant Agreement No. 101250730. Views and opinions expressed are however those of the author(s) only and do not necessarily reflect those of the European Union, EuroHPC Joint Undertaking or Poland. Neither the European Union, the EuroHPC Joint Undertaking nor Poland can be held responsible for them.
\end{acknowledgement}

\bibliography{agent_perspective}

\end{document}